\documentclass[usenatbib]{mn2e}
\usepackage{graphicx}

\def\eps@scaling{1.0}%
\newcommand\epsscale[1]{\gdef\eps@scaling{#1}}%
\newcommand\plotone[1]{%
 \centering
 \leavevmode
 \includegraphics[width={\eps@scaling\columnwidth}]{#1}%
}%
\newcommand\plottwo[2]{%
 \centering
 \leavevmode
 \columnwidth=1.0\columnwidth
 \includegraphics[width={\eps@scaling\columnwidth}]{#1}%
 \hfil
 \includegraphics[width={\eps@scaling\columnwidth}]{#2}%
}%

\newcommand\facility[1]{#1}

\newcommand\ion[2]{#1$\;${\small{#2}}}

\usepackage{url}

\newcommand{\tauGPeff}{\tau_{\rm GP}^{\rm eff}}
\newcommand{\xHI}{\bar{x}_{\rm HI}}

\newcommand{\avenf}{\xHI}
\newcommand{\lsim}{\la}

\newcommand{\lya}{Ly$\alpha$}
\newcommand{\lyb}{Ly$\beta$}

\newcommand{\HI}{\ion{H}{I}}

\newcommand{\myshorttitle}{Constraints on the $z\sim6$ neutral fraction}
\newcommand{\mytitle}{Model-independent evidence in favor of an end to reionization by $z\approx6$}

\begin{document}

\title[\myshorttitle]{\mytitle}

\author[I. D. McGreer, A. Mesinger, \& V. D'Odorico]{
Ian D. McGreer,$^{1,}$\thanks{Email: imcgreer@as.arizona.edu}
Andrei Mesinger,$^{2}$ and
Valentina D'Odorico$^{3}$\\
$^{1}$ Steward Observatory, The University of Arizona, 
                 933 North Cherry Avenue, Tucson, AZ 85721--0065 \\
$^{2}$ Scuola Normale Superiore, Piazza dei Cavalieri 7, I-56126 Pisa, Italy \\
$^{3}$ INAF--OATS, Via Tiepolo 11, I-34143 Trieste, Italy \\
}

\voffset-0.6in

\maketitle

\begin{abstract}
We present new upper limits on the volume-weighted neutral hydrogen fraction, 
$\avenf$, at $z\sim5\mbox{--}6$ derived from spectroscopy of bright quasars.
The fraction of the \lya\ and \lyb\ forests that is ``dark'' (with zero flux) 
provides the only model-independent upper limit on $\avenf$, requiring no 
assumptions about the physical conditions in the intergalactic medium or the 
quasar's unabsorbed UV continuum. In this work we update our previous results 
using a larger sample (22 objects) of medium-depth ($\sim$ few hours) spectra 
of high-redshift quasars obtained with the Magellan, MMT, and VLT. This 
significantly improves the upper bound on $\avenf$ derived from dark pixel 
analysis to $\avenf~\leq~0.06{+0.05~(1\sigma)}$ at $z=5.9$
and $\avenf~\leq~0.04{+0.05~(1\sigma)}$ at $z=5.6$. 
These results provide robust constraints for theoretical models of 
reionization, and provide the strongest available evidence that reionization 
has completed (or is very nearly complete) by $z\approx6$.
\end{abstract}


\section{Introduction}\label{sec:intro}

Cosmic reionization was the last major phase change of the baryons in our Universe.  The state of the intergalactic medium (IGM) during this epoch 
provides unique insight into the nature of the first structures. Thus 
observational probes of reionization have been long sought-after.
The first suggestive evidence of reionization came from the rapid redshift 
evolution of observed flux in the \lya\ and \lyb\ forests of $z\sim6$ QSO spectra 
(e.g. \citealt{Fan+01, White+03}).  Subsequent reionization studies have 
exploited: (i) the size and evolution of the proximity zone around quasars 
(e.g. \citealt{WLC05, Fan+06, Carilli10,BH07_quasars, Maselli07}); (ii) the 
damping wing absorption from neutral IGM in quasar and gamma-ray burst spectra 
(e.g. \citealt{MH04, Totani06, MH07, McQuinn08, Bolton11, SMH12, Totani+13}); 
(iii) the number density and clustering of Ly$\alpha$ emitters (e.g. 
\citealt{MR04, Kashikawa06, McQuinn07LAE, DMW11, DMF11, Jensen13, Konno14, 
Mesinger14}); (iv) the distribution of dark gaps in quasar spectra (e.g. 
\citealt{Croft98, SC02, GCF06, Gallerani08, Mesinger10}); (v) the integral 
constraints provided by the CMB primary (e.g. \citealt{Hinshaw13}) and 
secondary (e.g. \citealt{Zahn12, MMS12}) anisotropies.

Despite these efforts, no consensus has emerged as to whether or not we are 
witnessing the epoch of reionization at $z\sim6$. 
The many uncertainties and degeneracies inherent in modeling high-redshift 
astrophysics complicate the interpretation of the observational data. This 
problem is exacerbated by the expectation that reionization is very 
inhomogeneous on large-scales (e.g. \citealt{FZH04, Zahn11}). As such, 
the statistics available today are too limited to provide robust constraints; 
thus it is not even certain that reionization has completed by $z\lsim6$, 
as cosmic neutral patches of the IGM could easily masquerade as saturated 
regions of the Ly$\alpha$ forest \citep{Lidz07, Mesinger10}.

In a previous work \citep[][hereafter Paper I]{MMF11}, we examined the
Ly$\alpha$ and Ly$\beta$ forests of high-redshift quasars with spectra 
from the ESI instrument on the Keck II telescope. After binning the 
spectra to increase the dynamic range, we measured the fraction of 
binned pixels that were dark in both forests (the so-called 
``dark fraction''; \citealt{Mesinger10}). Using just two high 
signal-to-noise (S/N) spectra with $\sim10$~hr integration times, we 
constrained the volume-weighted mean neutral 
fraction of the IGM at $z\approx5.9$ to be $\avenf\lsim0.3$~($1\sigma$).
The dark fraction is only an upper limit on $\avenf$, as the residual \HI\  
inside the ionized IGM can also be sufficient to saturate \lya\ and \lyb\ at 
these redshifts (i.e. the dark patches need not correspond to cosmic \HI\  
regions during patchy reionization).  Nevertheless, our upper limits are 
competitive, and more importantly, are {\it the only model-independent 
limits on $\avenf$ to date.}

In this work we update our results from Paper I.  We apply the same analysis 
on a much larger sample of $z\sim6$ quasar spectra having total integration 
times of several hours on 6-10m class telescopes. We present the new 
observations from Magellan, MMT, and the VLT in \S\ref{sec:data}, then
summarize the dark pixel analysis of the new spectra in \S\ref{sec:analysis}.
The result is a much stronger upper limit on the neutral hydrogen fraction
at $z<6$ as detailed in \S\ref{sec:results}. 
We adopt a $\Lambda$CDM cosmology with identical parameters as in Paper I:
($\Omega_\Lambda$, $\Omega_{\rm M}$, $H_0$) = 
 (0.72, 0.28, 70 km s$^{-1}$ Mpc$^{-1}$; \citealt{Komatsu+09}); 
this is required only to convert redshift bins to comoving distances
and thus our results have negligible dependence on the choice of
cosmological model.

\section{Data}\label{sec:data}

In Paper I we presented results derived from 13 spectra obtained with the
ESI instrument on the Keck II telescope. With the exception of two deep
($\sim10$~hr) spectra, the data came from shallow ($\leq1$~hr) observations.
We include the original ESI sample in the analysis presented here, but add
new, deeper observations of $z\sim6$ quasars obtained with moderate resolution
spectrographs on the Magellan, MMT, and VLT telescopes. The complete list
of spectra is given in Table~\ref{tab:speclist}; the following sections 
describe the new observations.

\subsection{Magellan Spectra}\label{sec:magellan}

We observed four quasars at the Magellan Clay 6.5m on 2011 Jun 10-13 using 
the Magellan Echellette spectrograph \citep[MagE;][]{mage}. MagE combines
a grating with two cross-dispersing prisms to project 15 spectral orders onto
a 2048 $\times$ 1024 E2V CCD, providing wavelength coverage from 3000\AA\ to 
1.05$\mu$m. We employed the 0.7\arcsec$\times$10\arcsec\ slit, yielding a 
resolution of $R\sim5800$. The native dispersion varies from 
0.3~\AA~pixel$^{-1}$ at the blue end to 0.6~\AA~pixel$^{-1}$ at the red end.
Conditions were generally clear and non-photometric, with seeing ranging 
from 0\farcs6--0\farcs8.

\begin{figure*}
 \epsscale{2.05}
 \plotone{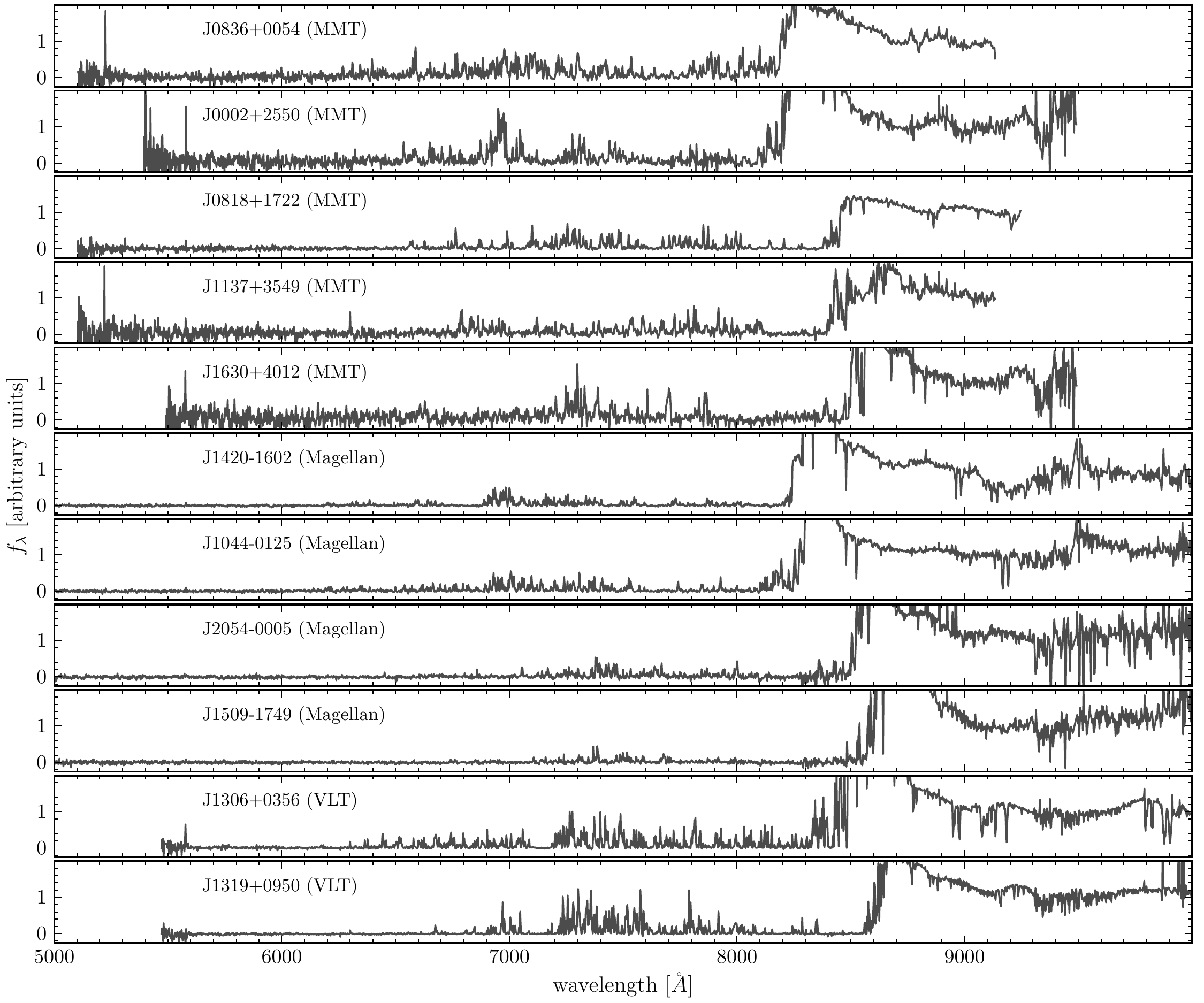}
 \caption{Spectra of $z\sim6$ quasars used for the dark pixel analysis. 
 Further details for each spectrum are given in Table~\ref{tab:speclist}. 
 The spectra have been rebinned to a dispersion of $\sim2.5$~\AA~pix$^{-1}$ 
 for clarity, and normalized by their flux density at rest-frame 
 $\sim$1350\AA; the displayed scale is arbitrary but the same in each
 panel. The transmission spikes become sharper as the resolution increases
 from the MMT to Magellan to VLT data.
  \label{fig:specfig}}
\end{figure*}

The Magellan spectra were processed with MASE \citep{mase}, an IDL-based
pipeline designed for MagE data. The calibration procedures outlined in 
\citet{mase} were followed. In brief, wavelength calibration is obtained 
from thorium-argon arcs observed immediately after the science targets. The
pixel flat-field images, slit illumination corrections, and the order traces 
are obtained from a combination of internal lamps and sky flats. 
The spectrophotometric standard star Feige 110 was observed each night to
provide flux calibration. With the exception of J1420-1602,
spectra for each target were obtained over multiple nights. The spectra for
each night were processed separately and then scaled and combined with 
inverse-variance weighting using the MASE routine \texttt{LONG\_COMBSPEC}. 
The variance does not include an estimate for the Poisson contribution due
to the object flux, as this would bias the result in low-count regions
\citep[see discussion in][]{White+03}.
The final spectra are displayed in Figure~\ref{fig:specfig}.

The MagE data from the first three nights are affected by increased and 
variable readnoise. The nominal readnoise is 3.1 e$^{-}$, which agrees with
measurements taken after the problem was addressed. Analysis of the overscan 
regions shows that the readnoise varied from $\sim$5--7 e$^{-}$ in images 
from the first $\sim2.5$ nights of observations. 
The dark pixel analysis we perform is concentrated on regions with low sky 
background between the strong night sky emission lines; hence the augmented 
readnoise diminishes the sensitivity of the spectra and has the 
effect of reducing the effective exposure time. We include readnoise estimates 
from the overscan regions in the variance calculation used by MASE in order
to account for this issue.

\begin{table}
 \begin{center}
 \caption{Quasar spectra\label{tab:speclist}}
 \begin{tabular}{lrrrrl}
 \hline
 Object & $z$ & $z_{\rm AB}$ & $t_{\rm exp}$ & $\langle\tau_{\rm eff,lim}^\alpha\rangle$ & source \\
 \hline
J1420-1602 &    5.73 &   19.7 &    4.00 &    5.3 & MagE \\
J0927+2001 &    5.77 &   19.9 &    0.33 &    3.8 & ESI \\
J1044-0125 &    5.78 &   19.2 &    4.79 &    5.2 & MagE \\
J0836+0054 &    5.81 &   18.7 &    0.33 &    4.7 & ESI \\
           &         &        &    4.00 &    3.9 & MMT \\
           &         &        &    2.27 &    5.9 & XShooter \\
J0002+2550 &    5.82 &   19.0 &    2.76 &    3.6 & MMT \\
J0840+5624 &    5.84 &   19.8 &    0.33 &    4.1 & ESI \\
J1335+3533 &    5.90 &   20.1 &    0.33 &    3.8 & ESI \\
J1411+1217 &    5.90 &   19.6 &    1.00 &    3.7 & ESI \\
J0148+0600 &    5.92 &   19.4 &   10.00 &    6.3 & XShooter \\
J0841+2905 &    5.98 &   19.8 &    0.33 &    3.5 & ESI \\
J1306+0356 &    6.02 &   19.5 &    0.25 &    4.3 & ESI \\
           &         &        &   11.50 &    5.4 & XShooter \\
J0818+1722 &    6.02 &   19.6 &    4.50 &    4.6 & MMT \\
           &         &        &    5.90 &    5.7 & XShooter \\
J1137+3549 &    6.03 &   19.5 &    0.67 &    3.8 & ESI \\
           &         &        &    2.33 &    3.7 & MMT \\
J2054-0005 &    6.04 &   20.7 &   11.00 &    4.4 & MagE \\
J0353+0104 &    6.05 &   20.5 &    1.00 &    3.5 & ESI \\
J1630+4012 &    6.07 &   20.4 &    4.39 &    3.2 & MMT \\
J0842+1218 &    6.08 &   19.6 &    0.67 &    4.0 & ESI \\
J1509-1749 &    6.12 &   20.3 &    6.00 &    4.7 & MagE \\
           &         &        &    8.32 &    5.2 & XShooter \\
J1319+0950 &    6.13 &   20.0 &   10.00 &    5.7 & XShooter \\
J1623+3112 &    6.25 &   20.1 &    1.00 &    4.2 & ESI \\
J1030+0524 &    6.31 &   20.0 &   10.32 &    5.3 & ESI \\
           &         &        &    7.46 &    5.4 & XShooter \\
J1148+5251 &    6.42 &   20.1 &   11.00 &    6.0 & ESI \\
 \hline
 \end{tabular}
 \end{center}
 Notes: Exposure times are given in hours.  
       $\langle\tau_{\rm eff,lim}^\alpha\rangle$ is the median effective optical depth in
       the \lya\ forest for a pixel (binned to 3.3 cMpc) with a flux equivalent to the $1\sigma$
       noise estimate.
\end{table}

\subsection{MMT Spectra}\label{sec:mmt}

Spectroscopy of five quasars was performed with the Red Channel spectrograph 
on the MMT 6.5m telescope during the nights of 2011 Jan 3-4 
(J0836+0054, J0818+1722, and J1137+3549)
and 2011 Jul 4-6 (J1630+4012 and J0002+2550). 
The Red Channel instrument includes an echellette mode that combines a grating 
with a single cross-disperser to provide complete coverage over a wavelength 
range of $\sim4000$\AA. We set the blaze angle of the grating to its nominal 
value of 4.555 and the cross-disperser at 350 (340 for the June observations). 
These  settings project roughly nine orders onto the 1024~$\times$~520 CCD, 
spanning $\sim5100$\AA\ to $\sim9000$\AA. The entrance aperture is a 
$1\arcsec\times10\arcsec$ slit and the resolution is nearly constant over the 
full wavelength range at $R\sim3300$ with a native dispersion of 
$\sim$0.6~\AA~pixel$^{-1}$. Both runs had typical seeing of $\sim1$\arcsec, but
were affected by clouds and thus varying transparency.

Compared to MagE, the Red Channel Echellette mode is less stable, 
and significant flexure affects exposures taken at different sky positions. 
In addition, observations were sometimes combined with longslit mode programs, 
necessitating reconfiguration of the echellette grating. Wavelength calibration 
is provided by HeNeAr arcs taken immediately after the science exposures and is
accurate to $<0.2$\AA\ (generally better than $0.1$\AA). On the other hand, 
typically only 1-2 quartz lamp exposures were obtained after the science 
exposures, and the nighttime images are not well aligned with the daytime 
flats. We thus combined all flats for each observing run into a single master 
flat and trimmed orders in individual images to match the coverage in the 
master flat. Target acquisition relied on offsets from nearby bright stars; 
we also obtained  brief spectra of the acquisition stars prior to applying 
the offset in order to aid with object tracing. Finally, flux calibration 
was obtained from observations of a single standard star each night.

The MMT spectra were processed with similar procedures as MASE using
custom-built routines implemented in Python. Bias subtraction was performed 
first using a low-order polynomial fit to the overscan region, and then
a second-order correction from a median-combined bias image. The master pixel
flats were then applied, and cosmic rays were identified and masked as outliers
in groups of images obtained for the same target in succession. A simple median
along image columns was used for sky subtraction as the slit axis is well
aligned with the image columns. Spectra of the acquisition reference stars 
were used to obtain the slit profiles of the objects. These profiles were then 
used for optimal extraction \citep{Horne86} of the 2D spectra. The extracted 
spectra were then flux calibrated, and the individual order spectra were then 
projected onto a common logarithmic wavelength grid and combined using 
inverse-variance weighting (again excluding the object contribution to the 
variance). As with the Magellan observations, nightly observations were 
processed independently and then combined to produce final spectra for each 
object, which are presented in Figure~\ref{fig:specfig}.

\subsection{VLT spectra}\label{sec:vlt}

The X-shooter instrument \citep{xshooter} on the VLT 8m Kueyen telescope 
was used to obtain spectra of 4 quasars at $z\sim6$ as part of a GTO 
program (P.I. V. D'Odorico) carried out between 2010 January and 2011
June. The spectra of three more quasars at $z\sim6$ observed at comparable
signal-to-noise ratio were downloaded from the X-shooter archive. The 
three arms of X-shooter allow for complete wavelength coverage between 
3000 \AA\ and 2.5 $\mu$m. The Ly$\alpha$ and Ly$\beta$ forests of the 
observed objects fall in the VIS arm, which spans the region between 
5500 \AA\ and 10300 \AA, approximately. The observations were carried out 
with a slit of width 0\farcs7 for J0836+0054, J1306+0356, J1319+0950, and J0148+0600; 
and of width 0\farcs9 for J0818+1722, J1030+0524 and J1509-1749; corresponding 
to resolutions $R\sim11000$ and $R\sim8800$, respectively.   
The raw spectra were reduced using the dedicated ESO pipeline
\citep{xspipe} until the science spectrum extraction step (see
below). A detailed description of the data reduction and the analysis
of metal absorption systems in six of the quasar spectra can be found in
\citet{DOdorico+13}.  

We found the pipeline-reduced X-shooter spectra had small positive residuals
in the dark portions of the spectra. This may be due to underestimation of
the mean sky level, or the method of combining spectra 
(see \S\ref{sec:magellan}). These residuals rose to the level of 
$\ga+0.2\sigma$ in some cases. While this would not affect most analyses, 
our dark pixel statistic is extremely sensitive to the pixel flux distribution
in zero-flux regions. Starting with the fully processed 2D spectra (i.e.,
after sky subtraction and rectification), we re-extracted the 1D spectra
using optimal extraction\footnote{The algorithm of \citep{Horne86} includes
the object contribution to the variance; when extracting our spectra -- 
including the MMT and Magellan spectra -- we remove this term to avoid the 
associated bias.} \citep{Horne86} and then combined the spectra 
without including the contribution from the object flux in the variance
\citep{White+03}. In a few cases, we also fit a smoothly varying spline 
function to the sky pixels in the 2D spectra prior to extraction, in order
to remove a small sky residual. After applying this post-processing, we 
found the pixel flux residuals after sky subtraction to be correctly 
distributed about zero for sky pixels. The reduced spectra of J1306+0356 
and J1319+0950 are shown in Figure~\ref{fig:specfig}.

\section{Dark Pixel Analysis}\label{sec:analysis}

We derive constraints on the neutral hydrogen fraction using an analysis of 
dark pixels in the spectra of high redshift quasars. Dark pixels are defined
by having zero measured flux; the motivation for using dark pixels as
a constraint on $\xHI$ is given in \citet{Mesinger10}. In brief, any physical
region containing pre-reionization neutral hydrogen will result in completely
saturated absorption in the resonant Lyman series transitions.  Therefore the
wavelength associated with that transition in the spectrum of a background 
quasar will have zero flux.
Dark pixels may also result from collapsed, self-shielded systems (e.g., 
Damped Lyman-alpha absorbers, or DLAs) and from post-reionization ionized 
gas with sufficient optical depth to render the flux undetectable given the 
finite $S/N$ limit of the spectrum.  Thus the dark fraction (fraction of the 
pixels which are dark) provides only an upper limit on $\avenf$.  
Nevertheless, this upper limit can be competitive with other estimates, is 
directly related to the volume filling factor of \HI, and {\it is completely 
independent of astrophysics}.  In contrast, the more common ``effective 
optical depth'' statistic, 
$\tauGPeff \equiv -\ln (\langle f_{\rm obs}/f_{\rm cont} \rangle_{\Delta z})$, 
is highly sensitive to: (i) transmission in rare voids (which is both difficult 
to simulate and not directly related to $\avenf$), and (ii) the estimate of 
the intrinsic quasar continuum, $f_{\rm cont}$ (e.g., \citealt{Lee12}).

In general we follow the same procedures outlined in Paper I to analyze the 
dark pixel fractions in our spectra, and we refer the reader to that work
for further details. For completeness we provide a brief summary of the 
methodology here, and discuss small changes to the analysis adopted for 
this work. 

First, for the purposes of our analysis, we use the term ``pixel'' to 
refer to a 3.3 comoving Mpc region obtained by binning the individual 
spectral pixels. Binning the data increases the dynamic range and thus 
strengthens our constraints, while the binned pixel size is chosen to retain 
information on physical structures at scales we are interested in (see 
Paper I for a more detailed justification of this choice). 
As we now have several instances of multiple spectra of the same quasar 
obtained from different telescope/instrument combinations, we combine the 
spectra at the stage of computing the binned pixels using inverse-variance
weighting of the individual contributions.
The dark pixel fractions are then calculated in wide redshift bins 
($\Delta{z} \sim 0.2$) from the binned pixels.
One modification to our analysis from Paper I is that we select
the redshift bins for the Ly$\alpha$ and Ly$\beta$ forests to avoid
wavelengths associated with regions of strong telluric absorption.

In Paper I we explored two methods for counting dark pixels in real 
(i.e., noisy) spectra. The first counted all pixels with $<2\sigma$ flux based 
on the rms  noise estimate in the spectrum, then scaled this number by the 
small fraction of zero flux pixels expected to scatter above this threshold 
(2.3\% assuming Gaussian statistics). The second method counted all pixels with 
a measured flux $<0$; in this case the scaling factor is two, as dark pixels 
are equally likely to have negative or positive values if the noise is symmetric 
around zero. The advantage of the first method is reduced uncertainties as more 
pixels contribute to the statistic. However, this method directly relies on 
the per-pixel noise estimate and thus a complete and precise accounting for 
all variance terms.  On the other hand, the negative pixel counts require only 
that the mean of the noise is zero and that the noise is symmetrically 
distributed about this value.

As a simple test of the pixel flux statistics for spectral regions without
any object flux, we examine the distribution of pixel fluxes for pixels
blueward of the Lyman Limit for each quasar spectrum. This region should be 
completely dark for $z\sim6$ quasars due to the high incidence of Lyman
Limit Systems (LLSs), and thus can be used to quantify any sky subtraction
residuals. We find that the pixels in our final processed spectra are indeed
symmetrically distributed about zero flux in the Lyman Limit region, although
the noise estimates in our heterogeneous dataset do not always strictly
follow Gaussian statistics\footnote{The ESI spectra used in Paper I, reduced 
using the method described in \citet{White+03}, do present highly Gaussian 
noise  properties according to this test.}.
We thus adopt only the negative pixel method in this work.

\begin{figure}
 \epsscale{1.0}
 \plotone{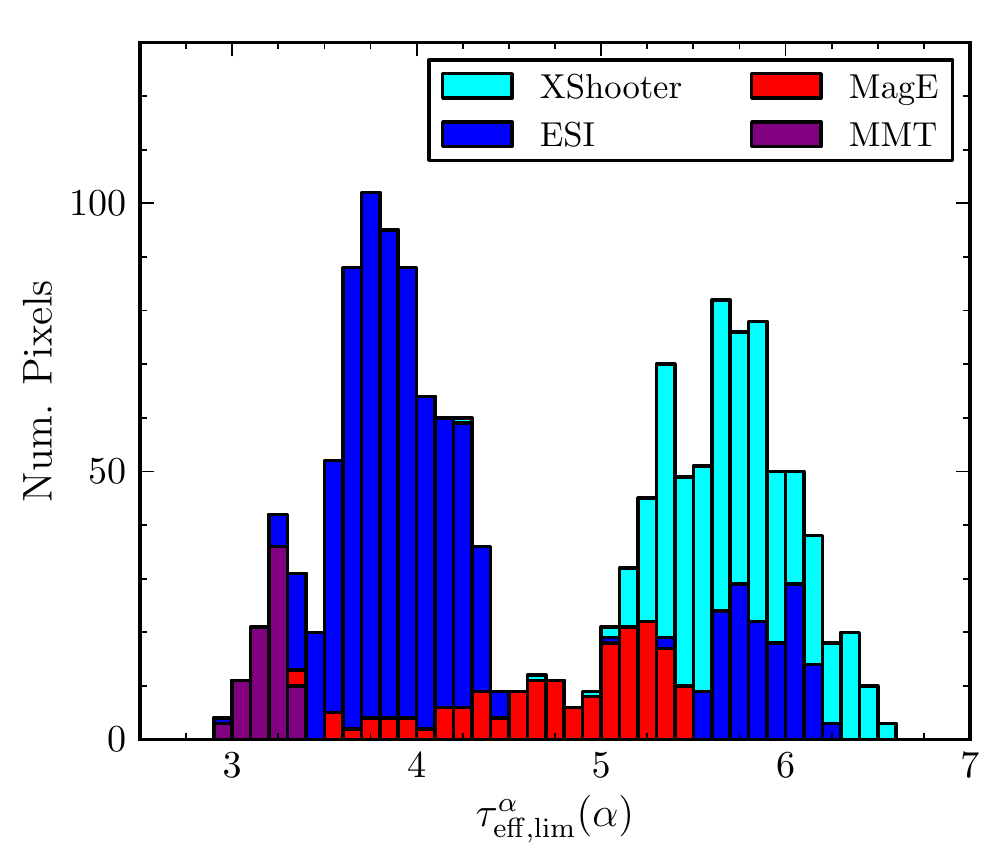}
 \caption{Depth of the Ly$\alpha$ pixels for the full dataset, quantified 
 by the $1\sigma$ limiting effective optical depth in a bin of 3.3 cMpc(see text for details). 
 The distribution for the Ly$\beta$ pixels is highly similar, although the
 numbers are smaller due to the shorter extent of the Ly$\beta$ forest.
 The ESI spectra used in Paper I were mainly shallow
 spectra with $\tau\la4$. This work represents a substantial increase in
 the number of pixels that probe depths of $\tau^{\alpha}_{\rm eff,lim} >4.5$. The pixels have
 been color-coded by the source of the spectrum; where multiple instruments
 were used the deeper spectrum is adopted (for example, two of the objects
 observed with MMT also have VLT observations).
 \label{fig:tauhist}}
\end{figure}

We use the metal line identifications compiled by \citet{DOdorico+13} 
and \citet{Becker+11} to mask wavelengths associated with strong
absorption systems. \citet{DOdorico+13} provide a catalog of absorption
systems associated with the high-ionization \ion{C}{IV} transition, while
the \citet{Becker+11} catalog includes low-ionization systems, namely
\ion{O}{I}, \ion{C}{II}, and \ion{Si}{II}. Although the actual amount of 
\HI\ absorption is unconstrained, we simply mask pixels at the Lyman series 
wavelengths associated with the metal line redshifts for all systems with
EW~$>0.3$~\AA. This includes masking Ly$\alpha$ absorption within the
Ly$\beta$ forest, as any DLA at these wavelengths would masquerade
as Ly$\beta$ dark pixels. In practice, few such systems fall within
our redshift bins (as expected from the incidence of DLAs at these
redshifts, e.g., \citealt{SC10}), so this has little effect on our
final results.

We quantify the depth of each pixel in the same manner as Paper I by
calculating a $1\sigma$ limit to the effective optical depth as
$\tau^{\rm eff}_{\rm lim} = -\ln(\langle\sigma/f_{\rm cont}\rangle)$, 
where $\sigma$ is the noise estimate for the binned pixels in the combined
spectra, and 
$f_{\rm cont}$ is the estimated continuum flux at the wavelength associated 
with the pixel\footnote{In Paper I we tabulated the $2\sigma$ limit, but 
since we are not using the $2\sigma$ threshold method to define dark pixels 
in this work we quote the $1\sigma$ limits.}. 
The continuum fit uses the broken power-law UV continuum model of 
\citet{Shull+12}, normalized to the flux at 1350~\AA.
Table~\ref{tab:speclist} lists the median limiting effective optical depths 
in the Ly$\alpha$ forests of each spectrum. We stress that these values 
merely serve to quantify the data quality, and are not used in deriving 
our results.

We use the same wavelength ranges for the Ly$\alpha$ 
and Ly$\beta$ forests: starting at
$1+z^\alpha_{\rm min} = (1+z_{\rm em})(1040 {\rm\AA}/\lambda_\alpha)$ and
$1+z^\beta_{\rm min} = (1+z_{\rm em})(970 {\rm\AA}/\lambda_\beta)$,
respectively, and ending at $z_{\rm QSO} - 0.1$ to avoid bias from the local
quasar overdensity. We calculate dark pixel fractions independently in the
two forests, and then in the combined \lya+\lyb\ forest by aligning the 
pixels in absorption redshift and then counting pixels that have negative 
flux in both forests. For the individual forests, the fraction of negative 
pixels are multiplied by two to obtain the $\xHI$\ constraint; this factor 
accounts for the additional pixels which intrinsically have zero flux, but 
are recorded to have positive values due to the noise.  Likewise, the 
fraction of pixels with recorded negative values in {\it both} \lya\ and 
\lyb\ are scaled by a factor of four to obtain $\avenf$ limits.  Again, this 
analysis assumes only a symmetric distribution around zero for zero-flux 
pixels. Finally, the uncertainties on the dark pixel fractions are 
calculated using the same jackknife method as described in Paper I; briefly, 
we account for sightline variance by computing the dark fractions in 
subsamples removing one quasar at a time, then use the resulting jackknife 
estimates of the mean and standard deviation from the subsamples for our 
analysis.

\begin{figure}
 \epsscale{1.0}
 \plotone{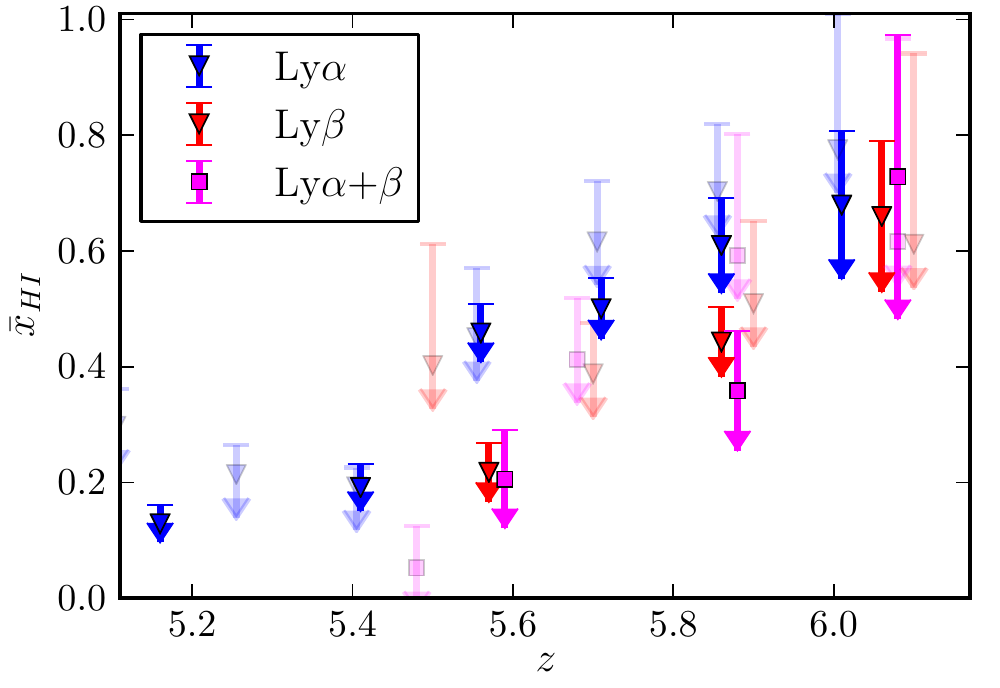}
 \caption{Comparison of the constraint on $\xHI$ obtained from the new 
 data (dark points) to the data used in Paper I (light points). There are
 small differences in the placement of the redshift bins, and the \lyb\ and
 \lya+\lyb\ points are offset slightly for clarity. Including 
 all of the data results in only a small improvement, although the error 
 bars are considerably reduced due to the better statistics.
 \label{fig:xhi_newdata}}
\end{figure}

\section{Results}\label{sec:results}

The new, deeper spectra presented here result in a significant improvement 
to our results from Paper I, in that many pixels that would previously
have been classified as ``dark'' are now detected in the higher $S/N$
spectra. These pixels sample regions of the IGM with large optical depths 
($4 \la \tau^\alpha_{\rm eff} \la 6$) but that are not associated with pre-reionization 
neutral regions. Figure~\ref{fig:tauhist} displays a histogram of the 
limiting effective optical depths of the spectra analyzed here. The new 
spectra, in particular from X-shooter, provide a much larger sample of 
high-$S/N$ pixels.
Figure~\ref{fig:xhi_newdata} compares our previous results directly, using
the same cut on spectral quality ($\tau^\alpha_{\rm eff,lim}>2.5$). It is apparent that the much
larger number of pixels greatly improves the statistics (as evident from the
smaller error bars), but has little effect on the resulting dark pixel 
fractions (and hence $\xHI$ constraint).

\begin{figure*}
 \epsscale{2.05}
 \plotone{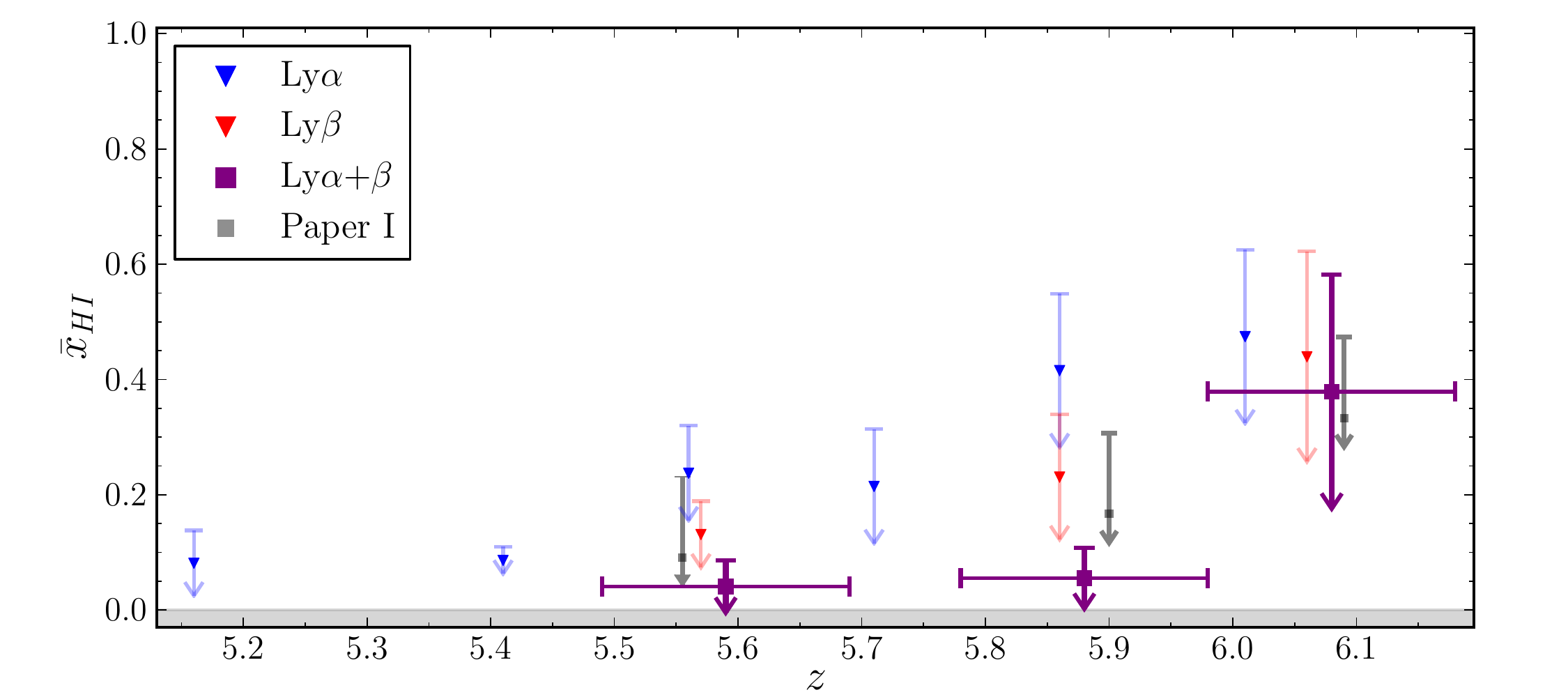}
 \caption{Constraint on the neutral hydrogen fraction from the best sample
 of spectra ($\tau^{\alpha}_{\rm eff,lim} > 4.5$). The individual \lya\ and \lyb\ 
 constraints are shown in light blue and light red, respectively, with
 vertical error bars representing the sightline-to-sightline scatter in 
 the dark fractions estimated from jackknife statistics. 
 The strongest constraints are obtained by combining 
 information from the \lya\ and \lyb\ forests, denoted by the dark purple 
 points, with horizontal error bars indicating the extent of the redshift 
 bins.
 The grey upper limits are from the two Keck ESI spectra presented in Paper I 
 (1$\sigma$ limits);  the $z\sim6.1$ constraint is not improved due to the
  lack of bright quasars at $z>6.2$, but now has model-independent error
  bars (see text for details).
 The constraints at $z<6$ are considerably stronger, with 
 $\xHI(\langle{z}\rangle=5.87) < 0.11$ and
 $\xHI(\langle{z}\rangle=5.58) < 0.09$ (both $1\sigma$).
  \label{fig:xhi2014}}
\end{figure*}

On the other hand, if we apply a much more stringent quality cut of 
$\tau^\alpha_{\rm eff,lim}>4.5$ using the new data, the resulting constraint is much stronger,
as seen in Figure~\ref{fig:xhi2014}. This cut is chosen to
reflect the bimodal distribution in $\tau$ seen in Fig.~\ref{fig:tauhist},
roughly dividing the spectra at one hour of integration time.
Applying this cut removes noisy pixels with insufficient sensitivity to probe
the large optical depth regions of the IGM. In Paper I, the deep sample
consisted of only two sightlines and was thus subject to cosmic variance,
whereas now each bin -- even in the combined Ly$\alpha$+Ly$\beta$ forest
with relatively small overlap -- contains at least four objects. 
We tabulate the dark pixel fractions, which translate directly into the 
$\xHI$ constraint, in Table~\ref{tab:data}. 

Introducing a quality cut using the effective optical depth measurements 
does tie our results to the continuum estimates, however, only very weakly. 
First, continuum errors on the order of 20--30\% will have little effect 
on the pixel distribution in our analysis (see Fig.~\ref{fig:tauhist}). 
Second, as the quasar continuum shape is uncorrelated with the absorption 
properties of the forest, continuum errors will cause pixels to move above 
or below the quality cut in an unbiased fashion. It is important to stress
that the quality cut only affects the selection of pixels to be included
in the analysis, the designation of dark pixels does not make use of the
continuum estimate. 

We find a single pixel with negative flux in the combined \lya+\lyb\ 
forests in each of the $z\approx5.6$ and $z\approx5.9$ redshift bins. 
As explained above, this translates to an expectation of $\sim4$ dark 
patches in each redshift bin, with the resulting $1\sigma$ jackknife 
limits of $\xHI <$ 0.11 (0.09) at $z\approx$ 5.9 (5.6).  These limits 
are a factor of $2-3$ improvement over the ones in Paper I, in which we 
were unable to conclude that reionization had completed  by $z\sim5$. With 
the available $S/N$ and an effective sightline length of $\sim$300 Mpc 
in each redshift bin, we expect $\sim$ few dark patches just from a 
post-reionization IGM (e.g. Fig. 6 in \citealt{Mesinger10}). Our results 
at  $z<6$ are therefore fully consistent with a post-overlap IGM.

Our limits of $\xHI <$ 0.11 (0.09) at $z\approx$ 5.9 (5.6) are the 
{\it most robust evidence to date that reionization has completed by 
$z\approx6$}.  In the context of simple patchy reionization models, this 
new limit on the end to reionization {\it is even more stringent than 
provided by the CMB Thompson scattering optical depth} 
(e.g. \citealt{MMS12}).

Our sample includes the quasar J0148+0600, which was reported by
\citet{Becker+14} to have a complete Gunn-Peterson trough in the \lya\ 
forest extending from $5.523~\le~z\le~5.879$. Indeed, this quasar
significantly increases the dark pixel fractions in the \lya\ forest at 
$z<6$. However, there are multiple transmission spikes in the \lyb\ forest.
In fact, the aligned  \lya\ and \lyb\ forests {\it have only a single pixel 
with negative flux in both forests} (corresponding to the value of four 
expected dark patches quoted in the above results for this redshift bin). 
The fact that the \lya\ forest is dark and yet the \lyb\ is not, implies 
that for a large part of the sightline the \lya\ optical depth is 
$6 \lsim \tau^\alpha_{\rm eff} \lsim 10$.  This is inconsistent with pre-overlap neutral 
patches (with $\tau \sim 10^5$).  Instead, this sightline demonstrates 
the presence of large-scale fluctuations in the photo-ionizing background 
at these redshifts \citep{Becker+14}, consistent with theoretical 
expectations (e.g. \citealt{Crociani11}).

\begin{table}
 \begin{center}
 \caption{Upper limits on $\xHI$ obtained from the high $S/N$ sample ($\tau^{\alpha}_{\rm eff,lim} > 4.5$)\label{tab:data}}
 \begin{tabular}{rrrrrr}
 \hline
  $z_1$ & $z_2$ & $N_{\rm LOS}$ & $N_{\rm pix}$ & $f_{\rm dark}$ & $+1\sigma$ \\
 \hline
\\[-8pt]
\multicolumn{5}{c} {\lya\ forest} \\
\\[-9pt]
   5.085 & 5.235 &   8 & 148 & 0.08 & 0.06 \\
   5.335 & 5.485 &  10 & 139 & 0.09 & 0.02 \\
   5.485 & 5.635 &  10 & 160 & 0.24 & 0.08 \\
   5.635 & 5.785 &   9 & 140 & 0.21 & 0.10 \\
   5.785 & 5.935 &   7 &  72 & 0.42 & 0.13 \\
   5.935 & 6.085 &   4 &  51 & 0.47 & 0.15 \\
\\[-8pt]
\multicolumn{5}{c} {\lyb\ forest} \\
\\[-9pt]
   5.480 & 5.680 &   6 & 104 & 0.13 & 0.06 \\
   5.770 & 5.970 &   6 &  96 & 0.23 & 0.11 \\
   5.970 & 6.170 &   4 &  50 & 0.44 & 0.18 \\
\\[-8pt]
\multicolumn{5}{c} {\lya+\lyb\ forest} \\
\\[-9pt]
   5.480 & 5.680 &   6 &  92 & 0.04 & 0.05 \\
   5.770 & 5.970 &   6 &  76 & 0.06 & 0.05 \\
   5.970 & 6.170 &   4 &  60 & 0.38 & 0.20 \\
 \hline
 \end{tabular}
 \end{center}
 Notes: The first two columns give the redshift ranges in
       which the dark pixel fractions are calculated. The third column is the
       number of lines-of-sight contributing to the redshift bin, and the 
       fourth column is the number of pixels in that bin. The fifth column is
       the dark pixel fraction (scaled from the negative pixel counts), and the
       final column is the $1\sigma$ uncertainty on this fraction from 
       jackknife statistics.
\end{table}

Finally, our highest redshift bin does not show improvement compared to 
the Paper I results, with $\xHI < 0.58$ at $z=6.1$~($1\sigma$).
We now have twice as many sightlines contributing to this bin
and thus employ the jackknife method to estimate the uncertainty (compared
to Paper I where we estimated the cosmic variance expected from only 
two sightlines using models of patchy reionization).\footnote{Our 4-6 
sightlines are able to sample the cosmic variance from patchy reionization 
to better than 1$\sigma$ for $\avenf \sim 0.1$ (see Fig. 6 in Paper I). 
We therefore consider our current sample size large enough to justify the 
use of jackknife statistics.} However, there is little
gain in depth, as the limiting $\tau^\alpha_{\rm eff}$ for the new data is similar to that
of the two deepest Keck spectra in Paper I. Thus the resulting constraint 
on $\avenf$ is not improved, with the the now model-independent 
error bars resulting in $1\sigma$ constraints somewhat weaker than our
previous estimate from two spectra. Further improvement in this redshift bin 
is limited by the lack of bright quasars with known redshifts greater than 
six; in addition, the Ly$\alpha$ forest is increasingly filled with strong 
night sky emission lines at wavelengths corresponding to this redshift, 
resulting in highly noisy spectra.

\section{Summary}\label{sec:summary}

We have updated our constraints on the neutral hydrogen fraction at 
$z\sim5\mbox{--}6$ using a sample of 20 high-S/N quasar spectra obtained 
with Magellan, MMT, and the VLT (complementing our previous work with 
Keck spectra). Using a simple, model-independent statistic --- the dark 
pixel fraction --- we obtain new $1\sigma$ upper limits of 
$\xHI < 0.11~(0.09)$ at $z=5.9~(5.5)$. The dark pixel statistic unambiguously 
demonstrates that reionization is highly complete by $z\la6$.
Our model-independent constraints can be used to confront theoretical models of 
inhomogeneous reionization.

Substantially improving on these new constraints would require an order of 
magnitude improvement in the statistics. Although Pan-STARRS is now 
delivering large samples of newly discovered bright quasars at $z\sim6$ 
\citep{Banados+14}, this is not likely enough to obtain a $<1\%$ constraint 
on $\xHI$ using dark pixel statistics. Therefore, we expect little 
observational improvements on our results.

On the other hand, the constraints on $\avenf$ can be improved by 
theoretical modeling of IGM properties and inhomogeneous reionization. 
Although by necessity this would involve model-dependencies, exploring a 
reasonably constrained set of IGM parameters would still yield reliable 
results that are consistent with the model-independent upper limits 
presented here. We leave this to a future work.

\subsection*{Acknowledgements}

We thank the referee for a careful read of the manuscript and suggestions
that improved the clarity of the text.
IDM acknowledges support from NSF grants AST 08-06861 and AST 11-07682.
Based in part on observations collected at the  European  Southern  Observatory  Very  Large  Telescope,  Cerro Paranal,  Chile -- Programs 084.A-0390, 084.A-0550, 085.A-0299, 086.A-0162, 087.A-0607 and 268.A-5767.
This paper also includes data obtained with the MMT Observatory, a joint 
facility  of the University of Arizona and the Smithsonian Institution,
and with the 6.5-m Magellan Telescopes located at Las Campanas Observatory, Chile.
The authors wish to recognize and acknowledge the very significant cultural role 
and reverence that the summit of Mauna Kea has always had within the indigenous 
Hawaiian community.  We are most fortunate to have the opportunity to conduct 
observations from this mountain.

{\it Facilities:}
 \facility{Keck:II (ESI)},
 \facility{MMT (Red Channel spectrograph)},
 \facility{Magellan:Clay (MagE)},
 \facility{VLT:Kueyen (X-shooter)}

\bibliographystyle{mn2e}
\bibliography{ms}

\begin{thebibliography}{}

\bibitem[\protect\citeauthoryear{{Ba{\~n}ados}, {Venemans}, {Morganson}
  et~al.,}{{Ba{\~n}ados} et~al.}{2014}]{Banados+14}
{Ba{\~n}ados} E.,  {Venemans} B.~P.,  {Morganson} E.,    et~al., 2014, \aj,
  148, 14

\bibitem[\protect\citeauthoryear{{Becker}, {Bolton}, {Madau}, {Pettini},
  {Ryan-Weber} \& {Venemans}}{{Becker} et~al.}{2014}]{Becker+14}
{Becker} G.~D.,  {Bolton} J.~S.,  {Madau} P.,  {Pettini} M.,  {Ryan-Weber}
  E.~V.,    {Venemans} B.~P.,  2014, ArXiv e-prints

\bibitem[\protect\citeauthoryear{Becker, Sargent, Rauch \& Carswell}{Becker
  et~al.}{2011}]{Becker+11}
Becker G.~D.,  Sargent W. L.~W.,  Rauch M.,    Carswell R.~F.,  2011, ApJ, 744,
  91

\bibitem[\protect\citeauthoryear{Bochanski, Hennawi, Simcoe et~al.,}{Bochanski
  et~al.}{2009}]{mase}
Bochanski J.~J.,  Hennawi J.~F.,  Simcoe R.~A.,    et~al., 2009, PASP, 121,
  1409

\bibitem[\protect\citeauthoryear{{Bolton} \& {Haehnelt}}{{Bolton} \&
  {Haehnelt}}{2007}]{BH07_quasars}
{Bolton} J.~S.,  {Haehnelt} M.~G.,  2007, \mnras, 374, 493

\bibitem[\protect\citeauthoryear{{Bolton}, {Haehnelt}, {Warren}, {Hewett},
  {Mortlock}, {Venemans}, {McMahon} \& {Simpson}}{{Bolton}
  et~al.}{2011}]{Bolton11}
{Bolton} J.~S.,  {Haehnelt} M.~G.,  {Warren} S.~J.,  {Hewett} P.~C.,
  {Mortlock} D.~J.,  {Venemans} B.~P.,  {McMahon} R.~G.,    {Simpson} C.,
  2011, \mnras, 416, L70

\bibitem[\protect\citeauthoryear{Carilli et~al.,}{Carilli
  et~al.}{2010}]{Carilli10}
Carilli C.~L.,  et~al., 2010, \apj, 714, 834

\bibitem[\protect\citeauthoryear{{Crociani}, {Mesinger}, {Moscardini} \&
  {Furlanetto}}{{Crociani} et~al.}{2011}]{Crociani11}
{Crociani} D.,  {Mesinger} A.,  {Moscardini} L.,    {Furlanetto} S.,  2011,
  \mnras, 411, 289

\bibitem[\protect\citeauthoryear{{Croft}}{{Croft}}{1998}]{Croft98}
{Croft} R.~A.~C.,  1998, in {A.~V.~Olinto, J.~A.~Frieman, \& D.~N.~Schramm}
  ed., Eighteenth Texas Symposium on Relativistic Astrophysics
  {Characterization of Lyman Alpha Spectra and Predictions of Structure
  Formation Models: A Flux Statistics Approach}.
pp 664--+

\bibitem[\protect\citeauthoryear{{Dayal}, {Maselli} \& {Ferrara}}{{Dayal}
  et~al.}{2011}]{DMF11}
{Dayal} P.,  {Maselli} A.,    {Ferrara} A.,  2011, \mnras, 410, 830

\bibitem[\protect\citeauthoryear{{Dijkstra}, {Mesinger} \& {Wyithe}}{{Dijkstra}
  et~al.}{2011}]{DMW11}
{Dijkstra} M.,  {Mesinger} A.,    {Wyithe} J.~S.~B.,  2011, \mnras, 414, 2139

\bibitem[\protect\citeauthoryear{D'Odorico, Cupani, Cristiani
  et~al.,}{D'Odorico et~al.}{2013}]{DOdorico+13}
D'Odorico V.,  Cupani G.,  Cristiani S.,    et~al., 2013, arXiv

\bibitem[\protect\citeauthoryear{Fan, Narayanan, Lupton et~al.,}{Fan
  et~al.}{2001}]{Fan+01}
Fan X.,  Narayanan V.~K.,  Lupton R.~H.,    et~al., 2001, AJ, 122, 2833

\bibitem[\protect\citeauthoryear{Fan, Strauss, Becker et~al.,}{Fan
  et~al.}{2006}]{Fan+06}
Fan X.,  Strauss M.~A.,  Becker R.~H.,    et~al., 2006, AJ, 132, 117

\bibitem[\protect\citeauthoryear{{Furlanetto}, {Zaldarriaga} \&
  {Hernquist}}{{Furlanetto} et~al.}{2004}]{FZH04}
{Furlanetto} S.~R.,  {Zaldarriaga} M.,    {Hernquist} L.,  2004, \apj, 613, 1

\bibitem[\protect\citeauthoryear{{Gallerani}, {Choudhury} \&
  {Ferrara}}{{Gallerani} et~al.}{2006}]{GCF06}
{Gallerani} S.,  {Choudhury} T.~R.,    {Ferrara} A.,  2006, \mnras, 370, 1401

\bibitem[\protect\citeauthoryear{{Gallerani}, {Ferrara}, {Fan} \&
  {Choudhury}}{{Gallerani} et~al.}{2008}]{Gallerani08}
{Gallerani} S.,  {Ferrara} A.,  {Fan} X.,    {Choudhury} T.~R.,  2008, \mnras,
  386, 359

\bibitem[\protect\citeauthoryear{{Hinshaw} et~al.,}{{Hinshaw}
  et~al.}{2013}]{Hinshaw13}
{Hinshaw} G.,  et~al., 2013, \apjs, 208, 19

\bibitem[\protect\citeauthoryear{Horne}{Horne}{1986}]{Horne86}
Horne K.,  1986, PASP, 98, 609

\bibitem[\protect\citeauthoryear{{Jensen}, {Laursen}, {Mellema}, {Iliev},
  {Sommer-Larsen} \& {Shapiro}}{{Jensen} et~al.}{2013}]{Jensen13}
{Jensen} H.,  {Laursen} P.,  {Mellema} G.,  {Iliev} I.~T.,  {Sommer-Larsen} J.,
     {Shapiro} P.~R.,  2013, \mnras, 428, 1366

\bibitem[\protect\citeauthoryear{Kashikawa et~al.,}{Kashikawa
  et~al.}{2006}]{Kashikawa06}
Kashikawa N.,  et~al., 2006, \apj, 648, 7

\bibitem[\protect\citeauthoryear{{Komatsu}, {Dunkley}, {Nolta}
  et~al.,}{{Komatsu} et~al.}{2009}]{Komatsu+09}
{Komatsu} E.,  {Dunkley} J.,  {Nolta} M.~R.,    et~al., 2009, \apjs, 180, 330

\bibitem[\protect\citeauthoryear{{Konno} et~al.,}{{Konno}
  et~al.}{2014}]{Konno14}
{Konno} A.,  et~al., 2014, ArXiv e-prints

\bibitem[\protect\citeauthoryear{{Lee}}{{Lee}}{2012}]{Lee12}
{Lee} K.-G.,  2012, \apj, 753, 136

\bibitem[\protect\citeauthoryear{{Lidz}, {McQuinn}, {Zaldarriaga}, {Hernquist}
  \& {Dutta}}{{Lidz} et~al.}{2007}]{Lidz07}
{Lidz} A.,  {McQuinn} M.,  {Zaldarriaga} M.,  {Hernquist} L.,    {Dutta} S.,
  2007, \apj, 670, 39

\bibitem[\protect\citeauthoryear{McGreer, Mesinger \& Fan}{McGreer
  et~al.}{2011}]{MMF11}
McGreer I.~D.,  Mesinger A.,    Fan X.,  2011, MNRAS, p.~1096

\bibitem[\protect\citeauthoryear{{Malhotra} \& {Rhoads}}{{Malhotra} \&
  {Rhoads}}{2004}]{MR04}
{Malhotra} S.,  {Rhoads} J.~E.,  2004, \apjl, 617, L5

\bibitem[\protect\citeauthoryear{Marshall, Burles, Thompson et~al.,}{Marshall
  et~al.}{2008}]{mage}
Marshall J.~L.,  Burles S.,  Thompson I.~B.,    et~al., 2008, in Ground-based
  and Airborne Instrumentation for Astronomy II {Proceedings of SPIE}.
SPIE, pp 701454--701454--10

\bibitem[\protect\citeauthoryear{{Maselli}, {Gallerani}, {Ferrara} \&
  {Choudhury}}{{Maselli} et~al.}{2007}]{Maselli07}
{Maselli} A.,  {Gallerani} S.,  {Ferrara} A.,    {Choudhury} T.~R.,  2007,
  \mnras, 376, L34

\bibitem[\protect\citeauthoryear{{McQuinn}, {Hernquist}, {Zaldarriaga} \&
  {Dutta}}{{McQuinn} et~al.}{2007}]{McQuinn07LAE}
{McQuinn} M.,  {Hernquist} L.,  {Zaldarriaga} M.,    {Dutta} S.,  2007, \mnras,
  381, 75

\bibitem[\protect\citeauthoryear{{McQuinn}, {Lidz}, {Zaldarriaga}, {Hernquist}
  \& {Dutta}}{{McQuinn} et~al.}{2008}]{McQuinn08}
{McQuinn} M.,  {Lidz} A.,  {Zaldarriaga} M.,  {Hernquist} L.,    {Dutta} S.,
  2008, \mnras, 388, 1101

\bibitem[\protect\citeauthoryear{Mesinger}{Mesinger}{2010}]{Mesinger10}
Mesinger A.,  2010, MNRAS, 407, 1328

\bibitem[\protect\citeauthoryear{{Mesinger}, {Aykutalp}, {Vanzella},
  {Pentericci}, {Ferrara} \& {Dijkstra}}{{Mesinger} et~al.}{2014}]{Mesinger14}
{Mesinger} A.,  {Aykutalp} A.,  {Vanzella} E.,  {Pentericci} L.,  {Ferrara} A.,
     {Dijkstra} M.,  2014, ArXiv e-prints:1406.6373

\bibitem[\protect\citeauthoryear{{Mesinger} \& {Haiman}}{{Mesinger} \&
  {Haiman}}{2004}]{MH04}
{Mesinger} A.,  {Haiman} Z.,  2004, \apjl, 611, L69

\bibitem[\protect\citeauthoryear{{Mesinger} \& {Haiman}}{{Mesinger} \&
  {Haiman}}{2007}]{MH07}
{Mesinger} A.,  {Haiman} Z.,  2007, \apj, 660, 923

\bibitem[\protect\citeauthoryear{{Mesinger}, {McQuinn} \& {Spergel}}{{Mesinger}
  et~al.}{2012}]{MMS12}
{Mesinger} A.,  {McQuinn} M.,    {Spergel} D.~N.,  2012, \mnras, 422, 1403

\bibitem[\protect\citeauthoryear{Modigliani, Goldoni, Royer et~al.,}{Modigliani
  et~al.}{2010}]{xspipe}
Modigliani A.,  Goldoni P.,  Royer F.,    et~al., 2010, in Silva D.~R.,  Peck
  A.~B.,   Soifer B.~T.,  eds, SPIE Astronomical Telescopes and
  Instrumentation: Observational Frontiers of Astronomy for the New Decade {The
  X-shooter pipeline}.
SPIE, pp 773728--773728--12

\bibitem[\protect\citeauthoryear{Schroeder, Mesinger \& Haiman}{Schroeder
  et~al.}{2012}]{SMH12}
Schroeder J.,  Mesinger A.,    Haiman Z.,  2012, MNRAS, 428, 3058

\bibitem[\protect\citeauthoryear{Shull, Stevans \& Danforth}{Shull
  et~al.}{2012}]{Shull+12}
Shull J.~M.,  Stevans M.,    Danforth C.~W.,  2012, ApJ, 752, 162

\bibitem[\protect\citeauthoryear{{Songaila} \& {Cowie}}{{Songaila} \&
  {Cowie}}{2002}]{SC02}
{Songaila} A.,  {Cowie} L.~L.,  2002, \aj, 123, 2183

\bibitem[\protect\citeauthoryear{Songaila \& Cowie}{Songaila \&
  Cowie}{2010}]{SC10}
Songaila A.,  Cowie L.~L.,  2010, ApJ, 721, 1448

\bibitem[\protect\citeauthoryear{Totani, Aoki, Hattori et~al.,}{Totani
  et~al.}{2013}]{Totani+13}
Totani T.,  Aoki K.,  Hattori T.,    et~al., 2013, arXiv:1312.3934v1

\bibitem[\protect\citeauthoryear{{Totani}, {Kawai}, {Kosugi}, {Aoki}, {Yamada},
  {Iye}, {Ohta} \& {Hattori}}{{Totani} et~al.}{2006}]{Totani06}
{Totani} T.,  {Kawai} N.,  {Kosugi} G.,  {Aoki} K.,  {Yamada} T.,  {Iye} M.,
  {Ohta} K.,    {Hattori} T.,  2006, \pasj, 58, 485

\bibitem[\protect\citeauthoryear{Vernet, Dekker, D'Odorico et~al.,}{Vernet
  et~al.}{2011}]{xshooter}
Vernet J.,  Dekker H.,  D'Odorico S.,    et~al., 2011, A \& A, 536, 105

\bibitem[\protect\citeauthoryear{White, Becker, Fan \& Strauss}{White
  et~al.}{2003}]{White+03}
White R.~L.,  Becker R.~H.,  Fan X.,    Strauss M.~A.,  2003, AJ, 126, 1

\bibitem[\protect\citeauthoryear{{Wyithe}, {Loeb} \& {Carilli}}{{Wyithe}
  et~al.}{2005}]{WLC05}
{Wyithe} J.~S.~B.,  {Loeb} A.,    {Carilli} C.,  2005, \apj, 628, 575

\bibitem[\protect\citeauthoryear{{Zahn} et~al.,}{{Zahn}  et~al.}{2012}]{Zahn12}
{Zahn} O.,  et~al., 2012, \apj, 756, 65

\bibitem[\protect\citeauthoryear{{Zahn}, {Mesinger}, {McQuinn}, {Trac}, {Cen}
  \& {Hernquist}}{{Zahn} et~al.}{2011}]{Zahn11}
{Zahn} O.,  {Mesinger} A.,  {McQuinn} M.,  {Trac} H.,  {Cen} R.,    {Hernquist}
  L.~E.,  2011, \mnras, 414, 727

\end{thebibliography}

\end{document}